\documentclass[aps,prl,amssymb]{revtex4}

\def\ep {\epsilon}

\def\e2 {\epsilon-\epsilon_k}
\def\be {\begin{equation}}
\def\ee {\end{equation}}
\def\bea {\begin{eqnarray}}
\def\eea {\end{eqnarray}}

\def\cd {c^{\dagger}}
\def\dd {d^{\dagger}}
\def\ua {\uparrow}
\def\da {\downarrow}
\def\si {\sigma}
\def\sp {\epsilon+U_p-5V}
\def\spu {\epsilon-U+U_p-5V}

\def\stu {\epsilon-U+U_p-4V}
\def\se  {\epsilon+U_p-6V}

\begin{document}
\title{An effective 1-band model for the cuprate superconductors }

\author{ George Kastrinakis}

\affiliation{ Institute of Electronic Structure and Laser (IESL), 
Foundation for Research and Technology - Hellas (FORTH),
P.O. Box 1527, Iraklio, Crete 71110, Greece$^*$}

\date{Dec. 31, 2008}

\begin{abstract}

Starting from the copper-oxygen Hamiltonian of the CuO$_2$ planes,
we derive analytically an extended 1-band Hubbard Hamiltonian for the 
electrons
on copper sites, through a canonical transformation which eliminates 
the oxygen sites.
The model sustains a variety of phases : checkerboard states, stripes,
antiferromagnetism, local pairs and mixtures thereof.
This approach may be helpful in understanding what is so special
about the CuO$_2$ planes, as opposed to other compounds. 

\end{abstract}

\maketitle

\vspace{.3cm}

{\bf I. INTRODUCTION }

\vspace{.3cm}

The starting point is the 3-band Hamiltonian, in which Cu
3$d_{x^2-y^2}$ and O 2$p_{x,y}$ orbitals are taken into account \cite{eme},
\be
H_o = \sum_{i,\si} \ep_i \dd_{i,\si} d_{i,\si} +
\sum_{i,j,\si} t_{oij} \dd_{i,\si} d_{j,\si} +
\sum_{i} U_i \dd_{i,\ua} d_{i,\ua} \dd_{i,\da} d_{i,\da} +
\sum_{i<j,\si, \si'} V_{ij} \dd_{i,\si} d_{i,\si} \dd_{j,\si'} d_{j,\si'} \;\;.
\ee
The creation/annihilation operators $\dd_{i,\si}/d_{i,\si}$ describe electrons
on the CuO$_2$ planes, and the indices $i,j$ run over {\em all} lattice sites.
The hopping matrix elements $t_{oij}=t,t'$ and the
off-diagonal Coulomb elements
$V_{i,j}=V,V'$ act between neighboring Cu and O atoms and between
neighboring O atoms respectively. $U_i=U,U_p$ and $\ep_i=\ep_d,\ep_p$
for Cu and O atoms.

The problem of reducing the 3-band Hamiltonian to a more amenable
effective 1-band Hamiltonian has been treated in a number
of papers \cite{zr,zo,er,rd,es,hy,bc,jef,kj}. Our goal is similar
in spirit. As the holes tend to reside mostly on the Cu atoms, we wish to
incorporate in the new effective Cu 1-band Hamiltonian explicit 2-particle
correlations, stemming from the original Hamiltonian. 

We emphasize that our approach is not a large-$U$ type (which was shown
to be problematic \cite{trem}), thus
allowing for double occupancy of the Cu sites. It merely eliminates the
oxygen sites. To this end, we use the canonical transformation method
of Chao, Spalek and Oles (CSO) \cite{cso}, 
adapted to the Hamiltonian of eq. (1) for the
CuO$_2$ planes. Using a Hamiltonian related, but not identical, to (1),
Zaanen and Oles (ZO) \cite{zo} applied this canonical transformation method 
to the cuprates.
Besides their different Hamiltonian, ZO followed a different strategy. The 
transformed Hamiltonian was separated into parts depending on the number 
of doubly 
occupied sites, and oxygen sites explicitly appeared therein, in contrast to 
our approach.

It is understood that the present method can also be applied to lattices
other than the CuO$_2$ plane.

In the following, section II contains the canonical transformation
formalism. In section III we present the new effective Hamiltonian $H$.
Section IV contains the solution for the ground state of $H$ along 
with a brief discussion on the phases encountered.

\vspace{.4cm}

{\bf II. CANONICAL TRANSFORMATION FORMALISM }

\vspace{.3cm}

Following CSO \cite{cso}, we write the Hamiltonian as 
\bea
H_o = H_A + H_B \;\;,  \label{hn}  \\ \nonumber
H_B = \sum_{i,j,\si} t_{oij} \dd_{i,\si} d_{j,\si} = 
\sum_{i\neq j} P_i H_o P_j \;\;, \\ \nonumber
H_A =  H_o - H_B = \sum_i P_i H_o P_i \;\;, 
\eea
with $P_i$ being projector operators with $\sum_j P_j = 1$. 
$P_i$ projects a state on the eigenstate
with eigenergy $E_i$ of the interacting part of $H_o$, i.e. $H_A|i>=E_i|i>$. 
Moreover, following 
CSO (c.f. before eq. (6a) of ref. \cite{cso}), we assume that 
\be
<i|H_B|j> = 0 \;\;,\;\; E_i = E_j \;\;, \label{enh1}
\ee
i.e. $H_B$ does {\em not} connect states which are {\em energetically 
degenerate}. This condition is further discussed below.

We consider the canonical transformation
\bea
H = e^{-iS} (H_A+H_B) e^{iS}  \label{exs} \\ \nonumber
= H_A + H_B - i[S,H_A] + \sum_{n=2}^{\infty} \frac{(-i)^n}{n!}
\left\{ [[S,H_A]]_n + in \; [[S,H_B]]_{n-1} \right\} \;\;,
\eea
where the operator $S$ is such that 
\be
H_B - i[S,H_A] = 0 \;\;. \label{exhs}
\ee
This condition amounts to the elimination of the O sites from $H$.
However, the matrix elements of $H$ between non-degenerate states
depend implicitly on the occupation of the O sites - c.f. below.
We use the notation
\be
[[A,B]]_n = [A,[A,[ ...,[A,B]]...]] \;\;,
\ee
with $n$ commutators at the right-hand side.

Substituting eq. (\ref{exhs}) into (\ref{exs}) yields
\be
H = H_A + \sum_{n=2}^{\infty} \frac{(n-1)(-i)^{n-1}}{n!} [[S,H_B]]_{n-1}
\;\; . \label{neah}
\ee
An expression for $S$ is derived by substituting into (\ref{exhs}) $H_A$
and $H_B$ from (\ref{hn}), and then apply the projectors $P_j$ from the left
and $P_k$ from the right on both sides of (\ref{exhs}), thus yielding
\be
P_j H_o P_k (1-\delta_{jk})+i P_j H_o P_j (P_j S P_k) 
- i (P_j S P_k) P_k H_o P_k = 0 \;\;. \label{ede}
\ee
Noting that $X_k=P_k H_o P_k = P_k H_A P_k$, $X_k$ are replaced by the proper 
energy eigenvalues $E_k$. For $E_j \neq E_k$ eq. (\ref{ede}) leads to
\be
P_j S P_k = i \frac{P_j H_B P_k}{E_j - E_k} \;\;.\label{xs1}
\ee
It also follows that \cite{cso}
\be
P_j S P_j = c \; P_j \;\;, \label{xs2}
\ee
with $c$ an arbitrary constant, whose value is irrelevant.
Then, using eqs. (\ref{xs1}), (\ref{xs2}) and $\sum_j P_j = 1$,
\bea
[[S,H_B]]_n = \left[ \left[ \sum_{j\neq k} P_j S P_k, -i \sum_{l \neq m}
(E_l - E_m) P_l S P_m \right] \right]_n  \\
= i(-1)^{n-1} \sum_{\text{all } k_m}'' \left\{ \sum_{j=0}^{n+1}
\frac{(-1)^j (n+1)!}{j! (n+1-j)!} E_{k_{j+1}} \right\} 
P_{k_1} S P_{k_2} S ... S P_{k_{n+2}} \;\;,
\eea
and the double primed summation is restricted to $E_{k_m} \neq E_{k_{m+1}}$ 
for all $m$, in accordance with condition (\ref{enh1}) above.
Substituting $[[S,H_B]]_n$ into eq. (\ref{neah}) yields
\be
H = H_A - \sum_{n=2}^{\infty} \frac{(n-1)i^n}{n!} 
\sum_{\text{all } k_m}'' \left\{ \sum_{j=0}^{n}
\frac{(-1)^j n!}{j! (n-j)!} E_{k_{j+1}} \right\} 
P_{k_1} S P_{k_2} S ... S P_{k_{n+1}} \;\;.
\ee
This is eq. (25) of CSO. Using eq. (\ref{xs1}), it can also be written as
\be
H = H_A - \sum_{n=2}^{\infty} \; (-1)^n \sum_{\text{all } k_m}'' 
I_n(\{k_m\}) \; P_{k_1} H_B P_{k_2} H_B ... H_B P_{k_{n+1}} \;\;,
\ee
where 
\be
I_n(\{k_m\}) = \frac{n-1}{n!} \; \sum_{j=0}^{n}
\frac{(-1)^j n!}{j! (n-j)!}  \;
\frac{ E_{k_{j+1}} }{\prod_{i=1}^n ( E_{k_{i}} - E_{k_{i+1}} )} \;\;. 
\label{ein}
\ee
Expressions for the factors $I_n$ used herein are given in Appendix A.

\vspace{.4cm}

{\bf III. EFFECTIVE HAMILTONIAN }

\vspace{.3cm}

Carrying out the expansion to fourth order in the hopping elements
$t,t'$, the new Hamiltonian turns out to be
\bea
H = \sum_{<l,j>,\si} \cd_{l,\si} c_{j,\si} \large\{ 
t_{1a} (1-n_{j,-\si}) (1-n_{l,-\si}) + t_{1b} 
[ n_{j,-\si}(1-n_{l,-\si}) + (1-n_{j,-\si}) n_{l,-\si} ] \\ \nonumber
+ t_{1c} n_{j,-\si} n_{l,-\si} +t_{1d} n_{j,-\si}(1-n_{l,-\si})
+t_{1e} (1-n_{j,-\si}) n_{l,-\si}   \large\} \\ \nonumber
-\sum_{<l,j;i>',\si} \cd_{l,\si} c_{j,\si} \Huge\{ 
(1-n_{i,-\si}) (1-n_{i,\si}) \Huge[ t_{2a} (1-n_{j,-\si}) 
(1-n_{l,-\si})  \\ \nonumber
+t_{2b} \{ n_{j,-\si}(1-n_{l,-\si}) + (1-n_{j,-\si}) n_{l,-\si} \}
+ t_{2c} n_{j,-\si} n_{l,-\si} \Huge]  
+t_{2d} n_{i,-\si} (1-n_{i,\si})
(1-n_{j,-\si}) n_{l,-\si}   \\ \nonumber
+ n_{i,\si} (1-n_{i,-\si}) \{ t_A (1-n_{l,-\si})(1-n_{j,-\si})
+t_B [n_{l,-\si}(1-n_{j,-\si}) + (1-n_{l,-\si}) n_{j,-\si} ]   
+ t_C n_{l,-\si} n_{j,-\si}  \}    \\ \nonumber
+ n_{i,\si} n_{i,-\si} \{ t_D (1-n_{l,-\si})(1-n_{j,-\si})
+t_E [n_{l,-\si}(1-n_{j,-\si}) + (1-n_{l,-\si}) n_{j,-\si} ] 
+ t_F n_{l,-\si} n_{j,-\si}  \} 
\Huge\}
\\ \nonumber
-\sum_{<l,j;i>,\si} \cd_{l,\si} c_{j,\si} \Huge\{ 
t_{2e} (1-n_{i,\si})(1-n_{i,-\si}) 
\\ \nonumber
+ t_{2f} \{ (1-n_{i,-\si}) n_{i,\si} + n_{i,-\si} (1-n_{i,\si}) \} 
+t_{2g} n_{i,-\si} n_{i,\si}  \Huge\} 
[ n_{j,-\si}(1-n_{l,-\si}) + (1-n_{j,-\si}) n_{l,-\si} ] \\ \nonumber
+U \sum_i n_{i,\ua} n_{i,\da}  
-A_{SXa} \sum_{<l,j>,\si} 
 n_{j,\si} n_{l,-\si} (1-n_{j,-\si}) (1-n_{l,\si})  \\ \nonumber
-\sum_{<l,j>,\si,\si'} \large\{ A_{SXb} \; n_{j,\si} n_{j,-\si}
[n_{l,\si'} (1-n_{l,-\si'})+n_{l,-\si'} (1-n_{l,\si'})] 
+ A_{SXc} \; n_{j,\si}(1- n_{j,-\si})
(1- n_{l,\si'})(1- n_{l,-\si'})  \large\}	\\ \nonumber
- A_{PT} \sum_{<l,j>,\si} \cd_{l,\si} \cd_{l,-\si} c_{j,-\si} c_{j,\si}
-A_{XE} \sum_{<l,j>,\si} \cd_{j,-\si} \cd_{l,\si} c_{j,\si} c_{l,-\si}
\\ \nonumber
- \sum_{<i,j;l>',\si} \Large\{ 
\cd_{l,\si} \cd_{l,-\si} c_{j,-\si} c_{i,\si}
+ \cd_{j,-\si} \cd_{i,\si} c_{l,\si} c_{l,-\si}  \Large\} \\ \nonumber
\Large\{ A_{FBa} \;(1-n_{j,\si}) (1-n_{i,-\si})  
+ A_{FBb}\; [ n_{j,\si}  (1-n_{i,-\si}) + (1-n_{j,\si}) n_{i,-\si} ]
+ A_{FBc}\; n_{j,\si} n_{i,-\si} \Large\}  \\ \nonumber
- \sum_{<i,j;l>'\si} \cd_{i,-\si} \cd_{l,\si} c_{j,\si} c_{l,-\si} \;
\large\{ A_{CMa} (1-n_{i,\si}) (1-n_{j,-\si})
+  A_{CMb} (1-n_{i,\si}) n_{j,-\si} \\ \nonumber
+ A_{CMc} n_{i,\si} (1-n_{j,-\si})
+ A_{CMd} n_{i,\si} n_{j,-\si} \large\}  \;\;.
\eea

The indices $i,j,l$ run exclusively over the Cu lattice and 
$n_{i,\si}=\cd_{i,\si} c_{i,\si}$. $<j,l>$ implies that $j$ and $l$
are nearest neighbors (n.n.), $<j,l;i>$ implies that $j$ and $l$
are second neighbors and $i$ is the common n.n. and $<j,l;i>'$
implies that in addition to second neighbors $j$ and $l$ can be third
neighbors with $i$ the middle common n.n.
The term 'empty lattice' below 
refers to the case of no other electrons present than the ones
hopping between initial and final positions; we ignore the electrons
in the rest of the lattice.
Henceforth $\ep=\ep_p-\ep_d$ and $V'=0$.

We see that the only term remaining intact from the original $H_o$
is the Hubbard term on Cu sites. All hopping terms now depend on the
site occupancy. The other new terms generated include superexchange
(SX), a pair-transfer (PT) between n.n., an exchange of electrons
(XE) between n.n., a local pair-formation or pair-breaking (FB), and
a correlated motion (CM) of two electrons. 
The respective matrix elements are given below. 


\vspace{.3cm}

{\bf Hopping elements}

\vspace{.3cm}

{\bf 1st neighbors} - all the terms of order $t^4$ below
contain a hopping forth and back between a Cu and an O atom.
Empty lattice, i.e. no other electron is present
in the Cu sites $j$ and $l$ involved
\bea
t_{1a} = -\frac{t^2}{\ep+U_p-6V}+\frac{4 t^2t'}{3} \left\{
\frac{1}{V(\ep+U_p-6V)}-\frac{1}{V(\ep+U_p-7V)}
-\frac{2}{(\ep+U_p-6V)(\ep+U_p-7V)} \right\} \\ \nonumber
 - \frac{3 t^2  {t'}^2}{2} \left\{ \frac{1}{V^2(\ep+U_p-7V)}
-\frac{1}{V(\ep+U_p-7V)^2} \right\} \\ \nonumber
+\frac{3t^4}{8(\ep+U_p-6V)(\ep+U_p-5V)} \left\{ \frac{1}{\ep-U+U_p-5V}
-\frac{3}{\ep-U+U_p-6V}  \right\}  \\  \nonumber
+ \frac{3t^4}{8(\ep-U+U_p-6V)(\ep-U+U_p-5V)} \left\{ \frac{1}{\ep+U_p-5V}
-\frac{3}{\ep+U_p-6V}  \right\}  \\  \nonumber
+ \frac{3t^4}{4(\ep+U_p-6V)(\ep-U+U_p-5V)^2} 
-\frac{9t^4}{4(\ep+U_p-6V)^2(\ep-U+U_p-5V)}   \\  \nonumber
+ \frac{3t^4}{(\ep+U_p-6V)(\ep+U_p-7V)^2}
+ \frac{3t^4}{(\ep+U_p-6V)^2(\ep+U_p-7V)}   \\  \nonumber
+ \frac{3t^4}{8(\ep-U+U_p-4V)(\ep+U_p-7V)} \left\{ 
\frac{1}{\ep+U_p-5V}+\frac{1}{\ep-U+U_p-5V}-\frac{6}{\ep+U_p-6V} \right\}
 \\  \nonumber
+\frac{3 t^4}{8(\ep+U_p-6V)}\left\{\frac{1}{\ep+U_p-5V}+\frac{1}{\ep-U+U_p-5V}
\right\} \left\{\frac{1}{\ep-U+U_p-4V}-\frac{3}{\ep+U_p-7V} \right\}
 \\  \nonumber
+\frac{3t^4}{2}\left\{ \frac{1}{(\ep+U_p-6V)^2 (\ep-U+U_p-6V)}
+\frac{1}{(\ep+U_p-6V) (\ep-U+U_p-6V)^2} \right\}
+\frac{2t^4}{(\ep+U_p-6V)^3}   \\  \nonumber
+\frac{3 t^4}{8 (\ep-U+U_p-4V)(\ep-U+U_p-5V)}
\left\{\frac{1}{\ep+U_p-6V}+\frac{1}{\ep+U_p-7V} \right\}   \\  \nonumber
-\frac{9 t^4}{8 (\ep+U_p-6V)(\ep+U_p-7V)}
\left\{ \frac{1}{\ep-U+U_p-4V}+\frac{1}{\ep-U+U_p-5V} \right\} \\  \nonumber
+\frac{3t^4}{4(\ep-U+U_p-5V)^2(\ep+U_p-6V)} 
-\frac{9t^4}{4(\ep-U+U_p-5V)(\ep+U_p-6V)^2} 
\;\; .
\eea

1st neighbors - an additional $-\si$ electron at either the initial site $j$
or final site $l$
\bea
t_{1b} = \frac{t^2}{2}\left\{ \frac{1}{\ep-U+U_p-5V} -\frac{1}{\ep+U_p-5V} 
\right\}
 - \frac{ t^2  {t'}^2}{8 V^2} \left\{ \frac{1}{\ep-U+U_p-6V}
+\frac{1}{\ep+U_p-7V}  \right\} 	\\ \nonumber
+ \frac{2 t^2t'}{3} \left\{
\frac{1}{2V(\ep-U+U_p-5V)}-\frac{1}{2V(\ep+U_p-7V)}
-\frac{2}{(\ep-U+U_p-5V)(\ep+U_p-7V)} \right\} \\ \nonumber
+ \frac{2 t^2t'}{3} \left\{
\frac{1}{V(\ep-U+U_p-6V)}-\frac{1}{V(\ep+U_p-5V)}
-\frac{2}{(\ep-U+U_p-6V)(\ep+U_p-5V)} \right\} 
\;\; .
\eea

1st neighbors - an additional $-\si$ electron at both initial
and final sites $j$ and $l$
\bea
t_{1c} = -\frac{t^2}{\ep-U+U_p-4V}  - \frac{3 t^2  {t'}^2}{8} 
\left\{ \frac{1}{V^2(\ep-U+U_p-6V)} -\frac{2}{V(\ep-U+U_p-6V)^2  } 
\right\} \\ \nonumber
+ \frac{ t^2t'}{3} \left\{
-\frac{1}{2V(\ep-U+U_p-6V)}+\frac{1}{2V(\ep-U+U_p-4V)}
-\frac{2}{(\ep-U+U_p-6V)(\ep-U+U_p-4V)} \right\}   \\ \nonumber
+\frac{t^4}{(\ep-U+U_p-4V)^3}
+\frac{3t^4}{2(\ep-U+U_p-6V)^2(\ep-U+U_p-4V)} \\ \nonumber
+\frac{3t^4}{2(\ep-U+U_p-6V)(\ep-U+U_p-4V)^2} 
\;\; . 
\eea

1st neighbors - an additional $-\si$ electron at the initial site j only
\bea
t_{1d} = \frac{3 t^4}{8 (\ep-U+U_p-4V)(\ep-U+U_p-5V)}
\left\{\frac{1}{\ep+U_p-5V}+\frac{1}{\ep-U+U_p-6V}\right\} \\ \nonumber
+\frac{3 t^4}{8(\ep-U+U_p-6V) (\ep+U_p-5V)}
\left\{ -\frac{3}{\ep-U+U_p-4V}+\frac{4}{\ep-U+U_p-5V} \right\} \\ \nonumber
+ \frac{3 t^4}{8 (\ep-U+U_p-5V)^2} \left\{\frac{1}{\ep-U+U_p-5V}
-\frac{5}{\ep+U_p-5V} \right\} \\ \nonumber
+\frac{3 t^4}{8(\ep-U+U_p-4V)(\ep+U_p-7V)} \left\{
\frac{1}{\ep-U+U_p-3V} -\frac{3}{\ep-U+U_p-6V} \right\} \\ \nonumber
+\frac{3 t^4}{8(\ep-U+U_p-3V)(\ep-U+U_p-6V)}  \left\{
\frac{1}{\ep-U+U_p-4V} -\frac{3}{\ep+U_p-7V}  \right\} \\ \nonumber
+\frac{3 t^4}{4(\ep+U_p-7V)(\ep-U+U_p-5V)} \left\{
-\frac{1}{\ep-U+U_p-3V}+\frac{2}{\ep+U_p-5V} \right\} \\ \nonumber
+\frac{3t^4}{8(\ep-U+U_p-5V)^2} \left\{
-\frac{3}{\ep+U_p-7V}+\frac{1}{\ep-U+U_p-3V}  \right\} \\ \nonumber
+\frac{3t^4}{8(\ep+U_p-7V)^2} \left\{
\frac{1}{\ep+U_p-5V}+\frac{3}{\ep-U+U_p-5V} \right\} \\ \nonumber
+\frac{3t^4}{8 (\ep-U+U_p-6V)^2} \left\{
\frac{1}{\ep-U+U_p-5V} -\frac{3}{\ep+U_p-5V} \right\} \\ \nonumber
+\frac{ t^4}{8 (\ep-U+U_p-5V)^2}\left\{
\frac{1}{\ep-U+U_p-5V}+\frac{7}{\ep+U_p-5V}    \right\}
+\frac{ 3t^4}{2 (\ep-U+U_p-5V)(\ep+U_p-5V) (\ep-U+U_p-6V)}  \\  \nonumber
+\frac{3t^4}{8 (\ep-U+U_p-6V)^2} \left\{ \frac{3}{\ep+U_p-5V}
+\frac{1}{\ep-U+U_p-5V} \right\}
\;\;.
\eea

1st neighbors - an additional $-\si$ electron at the final site $l$ only
\bea
t_{1e} = \frac{ t^4}{4(\ep-U+U_p-5V)^2}
\left\{ \frac{2}{\ep-U+U_p-5V}-\frac{7}{\ep+U_p-5V} \right\}  \\ \nonumber
+\frac{3t^4}{2(\ep-U+U_p-5V)(\ep+U_p-5V)(\ep-U+U_p-7V)} 
+\frac{3t^4}{(\ep-U+U_p-7V)^2} \left\{ \frac{3}{\ep-U+U_p-5V}
+\frac{1}{\ep+U_p-5V} \right\} \\ \nonumber
+\frac{3t^4}{8(\ep-U+U_p-5V)^2}\left\{ \frac{1}{\ep-U+U_p-3V}
-\frac{3}{\ep-U+U_p-7V} \right\}
-\frac{3t^4}{4(\ep-U+U_p-5V)(\ep-U+U_p-3V)(\ep+U_p-7V)}   \\ \nonumber
-\frac{9t^4}{8 (\ep-U+U_p-6V) (\ep+U_p-7V)} \left\{
\frac{1}{\ep-U+U_p-4V} + \frac{1}{\ep-U+U_p-3V} \right\}  \\ \nonumber
+\frac{3t^4}{8 (\ep-U+U_p-4V)(\ep-U+U_p-3V)} 
\left\{\frac{1}{\ep-U+U_p-6V}+\frac{1}{\ep+U_p-7V} \right\} \\ \nonumber
+\frac{3t^4}{8 (\ep-U+U_p-4V)(\ep-U+U_p-5V)} 
\left\{\frac{1}{\ep-U+U_p-6V}-\frac{1}{\ep+U_p-5V} \right\} \\ \nonumber
-\frac{9t^4}{8 (\ep-U+U_p-6V) (\ep+U_p-5V)} \left\{
\frac{1}{\ep-U+U_p-4V} + \frac{1}{\ep-U+U_p-5V} \right\}  \\ \nonumber
+\frac{3t^4}{8(\ep-U+U_p-6V)^2} \left\{\frac{1}{\ep-U+U_p-5V}
+\frac{3}{\ep+U_p-5V} \right\}
+\frac{3t^4}{2 (\ep-U+U_p-6V)(\ep-U+U_p-5V) (\ep+U_p-5V)}
\;\;.
\eea

{\bf 2nd/3rd neighbors} - empty lattice, i.e. no other electron is present
in the Cu sites involved, j initial, i intermediate and $l$ final 
\bea
t_{2a} = -\frac{t^4}{4} \left\{ \frac{1}{(\ep+U_p-5V)^2 (\ep+U_p-7V)}
- \frac{3}{(\ep+U_p-5V)(\ep+U_p-7V)^2} 
\right. \\ \nonumber
\left. 
+\frac{1}{(\ep+U_p-6V)^2 } \left[ \frac{1}{\ep+U_p-5V}
-\frac{3}{\ep+U_p-7V}\right]-\frac{6}{(\ep+U_p-6V)^3} 
\right\}  \;\;.
\eea

2nd/3rd neighbors - an additional $-\si$ electron at either the initial site j
or final site $l$
\bea 
t_{2b} =- \frac{t^4}{8 (\ep+U_p-7V) (\ep+U_p-4V) }
\left\{ \frac{1}{\ep+U_p-5V}+\frac{1}{\ep-U+U_p-5V}-\frac{3}{\ep-U+U_p-6V}
-\frac{3}{\ep+U_p-6V} \right\} \\ \nonumber
-\frac{t^4}{8} \left\{\frac{1}{\ep+U_p-4V} -\frac{3}{\ep+U_p-7V} \right\}
\left\{\frac{1}{(\ep+U_p-5V)(\ep-U+U_p-6V)} 
+\frac{1}{(\ep-U+U_p-5V)(\ep+U_p-6V)}  \right\} \\ \nonumber
-\frac{t^4}{8(\ep+U_p-5V)^2 } \left\{ \frac{1}{\ep+U_p-6V } 
+\frac{3}{\ep-U+U_p-6V} \right\}
-\frac{t^4}{4 (\ep+U_p-6V)^2 (\ep-U+U_p-5V) }  \\ \nonumber
+\frac{t^4}{4(\ep+U_p-5V) (\ep+U_p-6V)} 
\left\{ \frac{1}{\ep-U+U_p-6V } + \frac{2}{\ep-U+U_p-5V } \right\} 
 \;\;.
\eea

2nd/3rd neighbors - additional $-\si$ electron at both initial
and final sites j and $l$
\bea
t_{2c} =  \frac{t^4}{4(\ep+U_p-4V)^2 (\ep-U+U_p-6V)}
- \frac{3 t^4}{4(\ep+U_p-4V) (\ep-U+U_p-6V)^2}    \\ \nonumber
-\frac{3 t^4}{2(\ep+U_p-5V)^2 (\ep-U+U_p-5V) } 
+ \frac{t^4}{4 (\ep-U+U_p-5V) (\ep-U+U_p-6V)}
\left\{ \frac{1}{\ep+U_p-4V} -\frac{3}{\ep+U_p-5V} \right\}  \\ \nonumber
+\frac{t^4}{4 (\ep+U_p-4V)(\ep+U_p-5V)}
\left\{ \frac{1}{\ep-U+U_p-5V}-\frac{3}{\ep-U+U_p-6V} \right\}
\;\;.
\eea

2nd/3rd neighbors -  one additional $-\si$ electron at all 3 sites, initial j, 
final $l$ and intermediate i
\bea
t_{2d} = -\frac{t^4}{4} \left\{ \frac{1}{(\ep-U+U_p-5V)(\ep-U+U_p-3V)^2 }
- \frac{3}{(\ep-U+U_p-5V)^2(\ep-U+U_p-3V) } \right. \\ \nonumber
 \left. 
+\frac{1}{(\ep-U+U_p-3V)(\ep-U+U_p-4V)^2}  
-\frac{3}{(\ep-U+U_p-5V)(\ep-U+U_p-4V)^2 }+\frac{2}{(\ep-U+U_p-4V)^3}
\right\} \;\;. 
\eea

2nd/3rd neighbors -  one additional $\si$ electron at intermediate site i
\be
t_A = -\frac{t^4}{({\se})^3} \;\;.
\ee

2nd/3rd neighbors -  one additional $\si$ electron at intermediate site i
and one additional $-\si$ electron at either the initial site j or final 
site $l$
\be
t_B = -\frac{t^4}{8(\se)^2} \left\{ \frac{1}{\sp} + \frac{3}{\spu} \right\}
- \frac{t^4}{2(\se)(\sp)(\spu)} 
\;\;.
\ee

2nd/3rd neighbors -  one additional $\si$ electron at intermediate site i
and two additional $-\si$ electrons at both initial site j and final 
site $l$
\be
t_C = -\frac{3t^4}{2(\spu)^2(\sp)} - \frac{t^4}{4(\sp)^2(\spu)} \;\;.
\ee

2nd/3rd neighbors - two additional electrons at intermediate site i
\be
t_D = -\frac{t^4}{4(\spu)^2(\sp)} - \frac{3t^4}{4(\sp)^2(\spu)} \;\;.
\ee

2nd/3rd neighbors - two additional electrons at intermediate site i
and one additional $-\si$ electron at either the initial site j or final 
site $l$
\be
t_E = -\frac{t^4}{2(\sp)(\spu)(\stu)} 
- \frac{t^4}{8(\stu)^2} \left\{ \frac{3}{\sp} + \frac{1}{\spu} \right\} \;\;.
\ee

2nd/3rd neighbors - two additional electrons at intermediate site i
and two additional $-\si$ electrons at both initial site j and final 
site $l$
\be
t_F = - \frac{t^4}{(\stu)^3} 
\;\;.
\ee

{\bf 2nd neighbors only} - an additional $-\si$ electron at either 
the initial site j or final site $l$
\bea 
t_{2e} = \frac{ t^2 t'}{3} \left\{ \frac{1}{V(\ep-U+U_p-6V)}
-\frac{1}{V(\ep+U_p-7V)}- \frac{2}{(\ep-U+U_p-6V) (\ep+U_p-7V)}
\right\}    \;\;.
\eea

2nd neighbors only - one additional $-\si$ electron 
at either the initial site j
or final site $l$ and one at the intermediate site i
\bea
t_{2f} =\frac{ t^2 t'}{3} \left\{ \frac{1}{V(\ep+U_p-6V)}
-\frac{1}{V(\ep-U+U_p-5V)} + \frac{2}{(\ep+U_p-6V) (\ep-U+U_p-5V)}
\right\}
\;\;. 
\eea 

2nd neighbors only - one additional $-\si$ electron 
at either the initial site j
or final site $l$ and two at the intermediate site i
\bea
t_{2g} =\frac{  t^2 t'}{3} \left\{ \frac{1}{V(\ep+U_p-5V)}
-\frac{1}{V(\ep-U+U_p-4V)} + \frac{2}{(\ep+U_p-5V) (\ep-U+U_p-4V)}
\right\}
\;\;. 
\eea

\vspace{.3cm}

{\bf Other elements}

\vspace{.3cm}

{\bf Transfer of a pair to a nearest neighbor site}
\bea
A_{PT}=\frac{t^4}{2} \left\{\frac{1}{(\ep-U+U_p-5V)^2(\ep+U_p-5V)}
+ \frac{3}{(\ep-U+U_p-5V)(\ep+U_p-5V)^2}  \right\}  \;\; .
\eea

Transfer of a pair to a second neighbor site =$O(t^4 t'^2)$.

{\bf Formation/breaking of a pair} - the pair is at site $l$, 
no other electrons at final (pair breaking)/initial (pair formation) 
sites j and i  
\bea
A_{FBa} =  \frac{t^4 }{(\ep+U_p-5V)(\ep-U+U_p-5V)(\ep+U_p-6V) } \\ \nonumber
+ \frac{t^4}{4(\ep+U_p-6V)^2} \left\{ \frac{3}{\ep+U_p-5V}
- \frac{1}{\ep-U+U_p-5V} \right\} 
 \;\;.
\eea

Formation/breaking of a pair -  with a minus spin electron
either at site i or at site j
\bea
A_{FBb} =  \frac{t^4 }{2} \left\{ \frac{1}{(\ep+U_p-6V)^2 (\ep-U+U_p-4V)}
+ \frac{1}{(\ep+U_p-6V)(\ep-U+U_p-4V)^2 } \right\}  \\ \nonumber
- \frac{t^4 }{8 (\ep-U+U_p-3V) (\ep+U_p-5V)} \left\{ 
\frac{1}{\ep-U+U_p-4V } + \frac{1}{\ep+U_p-6V} \right\} \\ \nonumber
- \frac{t^4 }{4(\ep+U_p-5V)^2 (\ep-U+U_p-4V) }
- \frac{t^4 }{4(\ep+U_p-5V) (\ep-U+U_p-4V)^2 }  
 \;\; .
\eea

Formation/breaking of a pair - with minus spin electrons
at both sites i and j
\bea
A_{FBc} =  \frac{t^4 }{4(\ep-U+U_p-5V)^2} \left\{ \frac{1}{\ep-U+U_p-5V}
+\frac{5}{\ep+U_p-5V} \right\}  
\;\; .
\eea

{\bf Exchange of two electrons with opposite spin - nearest neighbor case}
\bea
A_{XE} =  \frac{t^4}{2 (\ep+U_p-5V)^2 (\ep-U+U_p-5V) }
+\frac{3 t^4}{2 (\ep+U_p-5V)(\ep-U+U_p-5V)^2  }   \\ \nonumber
+\frac{ t^4 t'}{ 30 \;V (\ep-U+U_p-6V)(\ep+U_p-5V) }
\left\{\frac{1}{\ep+U_p-5V} + \frac{10}{\ep-U+U_p-5V} \right\}  \\ \nonumber
+\frac{ t^4 t'}{30 (\ep+U_p-5V)^2 (\ep-U+U_p-5V) }
\left\{\frac{4}{\ep-U+U_p-6V} - \frac{1}{V} \right\} 
\;\;.
\eea

Exchange of two electrons with opposite spin - second neighbor case
= $O(t^4 t'^2)$.

{\bf Two electrons moving to neighboring sites} 
- empty lattice case
\bea
A_{CMa} = \frac{t^4}{4 U (\ep-U+U_p-5V)(\ep+U_p-5V)}
+\frac{3 t^4}{8 (\ep-U+U_p-5V)^2 } \left\{\frac{1}{U}+\frac{1}{\ep+U_p-5V}
\right\}  \\ \nonumber
-\frac{ t^4}{4 (\ep-U+U_p-5V)(\ep+U_p-5V)^2 } 
+\frac{3 t^4}{8 (\ep-U+U_p-5V)^2(\ep+U_p-5V)}  \\ \nonumber
-\frac{ t^4}{4 (\ep-U+U_p-4V)^2(\ep+U_p-6V)} 
+\frac{3 t^4}{4 (\ep-U+U_p-4V)(\ep+U_p-6V)^2}
+\frac{ t^4}{4 (\ep+U_p-6V)^3}   \\ \nonumber
+\frac{ 3t^4}{4 (\ep+U_p-6V)(\ep+U_p-5V) (\ep-U+U_p-5V) }  \\ \nonumber
+\frac{t^4}{ 4 (\ep-U+U_p-5V) (\ep-U+U_p-4V)} \left\{ \frac{3}{\ep+U_p-6V}
-\frac{1}{\ep+U_p-5V} \right\}
 \;\;.
\eea

Two electrons moving to neighboring sites 
- with a minus spin electron at site j
\bea
A_{CMb} = \frac{ t^4}{8 (\ep-U+U_p-4V)^2} \left\{ -\frac{1}{\ep+U_p-5V}
+\frac{9}{\ep-U+U_p-5V} \right\}   \\ \nonumber
-\frac{ t^4}{8  (\ep+U_p-6V)^2} \left\{ \frac{1}{\ep+U_p-5V}
+\frac{ 3}{\ep-U+U_p-5V}  \right\}     \\ \nonumber
+\frac{ t^4}{4 (\ep+U_p-5V) (\ep-U+U_p-5V)}
\left\{ \frac{1}{ \ep+U_p-6V } +\frac{4}{ \ep-U+U_p-4V } \right\}
\;\; .
\eea

Two electrons moving to neighboring sites 
- with a minus spin
electron at site i
\bea
A_{CMc} = \frac{  t^4}{8 (\ep-U+U_p-4V)^2 }    
\left\{ -\frac{1}{\ep+U_p-5V}+\frac{3}{\ep-U+U_p-5V} \right\} \\ \nonumber
+\frac{  t^4}{2 (\ep+U_p-5V)(\ep-U+U_p-5V)(\ep-U+U_p-4V) }
+\frac{  t^4}{8 (\ep-U+U_p-5V)^2} \left\{ \frac{1}{\ep-U+U_p-6V}
+ \frac{3}{\ep+U_p-6V} \right\}		\\ \nonumber
+\frac{ t^4}{ 4 (\ep+U_p-6V) (\ep-U+U_p-3V)} 
\left\{ \frac{3}{\ep-U+U_p-5V} - \frac{1}{\ep-U+U_p-4V} \right\}  \\ \nonumber
+\frac{t^4}{4 (\ep-U+U_p-4V) (\ep-U+U_p-5V)} 
\left\{- \frac{1}{\ep-U+U_p-3V} + \frac{3}{\ep+U_p-6V} \right\}  
\;\; .
\eea

Two electrons moving to neighboring sites 
- with minus spin
electrons at both sites $i$ and $j$
\bea
A_{CMd} = \frac{3 t^4}{2(\ep-U+U_p-4V)^3} 
+\frac{t^4}{4(\ep+U_p-5V)^2  (\ep-U+U_p-5V)}
+\frac{3 t^4}{4(\ep+U_p-5V)  (\ep-U+U_p-5V)^2}	\\ \nonumber
-\frac{t^4}{4 (\ep-U+U_p-3V)^2 (\ep-U+U_p-5V) } 
+\frac{3 t^4}{4 (\ep-U+U_p-3V) (\ep-U+U_p-5V)^2 }    \;\;.
\eea

{\bf Superexchange between two Cu sites } - one electron at site $j$
and a minus spin electron at site $l$
\bea
A_{SXa} = -\frac{t^4}{4 (\ep-U+U_p-5V) } 
\left\{\frac{1}{(\ep+U_p-5V)^2} -\frac{1}{(\ep-U-3V)^2 } \right\} \\ \nonumber
-\frac{3 t^4}{4 (\ep-U+U_p-5V)^2 } 
\left\{\frac{1}{\ep+U_p-5V} -\frac{1}{\ep-U-3V } \right\} 
\eea

Superexchange between two Cu sites  - two electrons at site $j$
and one electron at site $l$
\be
A_{SXb} = -\frac{t^4}{(\ep+U_p-4V)^3} \;\;.
\ee

Superexchange between two Cu sites  - one electron at site j only,
and no electron at site l, i.e. a renormalization of the site j energy,
due to site $l$ being empty
\be
A_{SXc} = -\frac{t^4}{(\ep+U_p-6V)^3} \;\;.
\ee

Typically \cite{hy}
\be
\ep=3.6 eV, t=1.3 eV, t'=0.65 eV, U=10.5 eV, U_p=4 eV, V=1.2 eV, V'=0 \;\;.
\ee

We emphasize that higher order terms are, in principle, of similar
magnitude as the terms shown. Energy level degeneracies, due to finite 
O-O hopping, appear in fifth order of perturbation theory, 
restricting the present
formulation. Similar issues arose in the original work
of CSO \cite{cso}. One should come up with a modified procedure, possibly
including an energy diagonalization in the vicinity of the Cu atom, 
as in \cite{jef}.
Of course it is possible that the series generated
are asymptotic anyway. Then the coefficients of the terms shown can be taken
as merely effective parameters.

It is interesting that for parameter values close to the "typical" ones,
factors such as
\be
\ep+U_p-m \; V\;, \;\;\; m=5,6 \;\;,
\ee
may become very {\em small} in magnitude, which yields increased values of the
respective interaction amplitudes $A$. It turns out that some
effective hopping elements increase at least equally fast in that case, 
so that the ratios
$(A/t)_{eff}$ are finite. However, this picture may be helpful
in understanding why certain e.g. 2-particle processes are important 
in the CuO$_2$ planes, as opposed to other lattices with different
values of the original parameters. Otherwise put, {\em what is so special
about the CuO$_2$ planes}.

\vspace{.4cm}

{\bf IV. GROUND STATE OF THE HAMILTONIAN }

\vspace{.3cm}

We can treat $H$ in the Hartree-Fock-Bogoliubov approximation, with the 
expectation values of four operator products given by
\be
<c_1 c_2 c_3 c_4>
=d_{12} d_{34} -d_{13}d_{24}+ d_{14}d_{23} \;\;,\;
\ee
where $d_{ij}=<c_{i,\sigma_i} c_{j,\sigma_j}>$ are numbers.
Yet another obvious approximation is the replacement of the operators
$n_{i\si}$ by their expectation values (else we would encounter 
expectaion values of six, instead of four, operator products).

In order to find the ground state numerically, we minimize $H$
with a fixed total number of particles $N=\sum_{i,\si}  n_{i,\si}$.
This procedure requires a highly sophisticated optimization solver,
able to handle several thousands of variables, with adequate
constraints on their values; overall a non-trivial task \cite{opt}.
In our implementation, we only looked at non-magnetized solutions.

As a first approach, we make one further simplification, taking
every fermion operator as a complex number (thus having only 4 real numbers
per lattice site - c.f. below).
In short, within this approach, we obtained checkerboard states with periods
equal to 3 by 3 and 5 by 5 (not 4) lattice sites, stripes, 
pure antiferromagnetic
states, local pairs and mixtures thereof. E.g. an $x-y$ anisotropic
checkerboard state is a mixture with a stripe state. These were found
for filling factors $n=N/V=0.8-1.2$ ($V$ the system volume). 
The nature of the ground state 
is mostly determined through the values of $U$ and the effective interaction
and hopping parameters, rather than the filling (of course the latter
dictates the values of the original CuO$_2$ plane parameters).

For a more complete solution, we should take all $d_{ij}$ above as
independent parameters. This amounts to 72 real numbers per lattice
site (with symmetry effects taken into consideration), 
making the problem very demanding computationally. 
The presentation of further results is postponed for a future
version of this work.


Yet another route to the ground state is through the new exact
variational wavefunctions which sustain superfluidity \cite{nf}.

\vspace{.3cm}

The author is indebted to Gregory Psaltakis for numerous discussions.

\vspace{.4cm}

{\bf APPENDIX A}
\vspace{.3cm}

Here we give explicit expressions for the first few factors $I_n$ of
eq. (\ref{ein})
\bea
I_2 = \frac{1}{2}\left(-\frac{1}{E_{12}}+\frac{1}{E_{23}} \right)\;\;, \\
I_3 = \frac{1}{3}\left(\frac{1}{E_{12}E_{23}} -\frac{2}{E_{12}E_{34}}
+ \frac{1}{E_{23}E_{34}} \right) \;\; , \\
I_4 = \frac{1}{8}\left(-\frac{1}{E_{12}E_{23}E_{34}}
+ \frac{3}{E_{12}E_{23}E_{45}}
-\frac{3}{E_{12}E_{34}E_{45}} + \frac{1}{E_{23}E_{34}E_{45}} \right) \;\; , \\
I_5 = \frac{1}{30}\left(\frac{1}{E_{12}E_{23}E_{34}E_{45}} 
- \frac{4}{E_{12}E_{23}E_{34}E_{56}} 
+ \frac{6}{E_{12}E_{23}E_{45}E_{56}} -\frac{4}{E_{12}E_{34}E_{45}E_{56}}
+\frac{1}{E_{23}E_{34}E_{45}E_{56}}   \right) \;\; ,  \\
I_6 = \frac{1}{144}\left(-\frac{1}{E_{12}E_{23}E_{34}E_{45}E_{56}} 
+\frac{5}{E_{12}E_{23}E_{34}E_{45}E_{67}}
-\frac{10}{E_{12}E_{23}E_{34}E_{56}E_{67}}
+\frac{10}{E_{12}E_{23}E_{45}E_{56}E_{67}} \right. \\ \nonumber
\left. -\frac{5}{E_{12}E_{34}E_{45}E_{56}E_{67}}
+\frac{1}{E_{23}E_{34}E_{45}E_{56}E_{67}}   \right) \;\;.
\eea
where $E_{ij} = E_{k_i} - E_{k_j}$.

\vspace{.3cm}
$^*$ e-mail : kast@iesl.forth.gr

\end{document}